\documentclass{ws-mpla}
\usepackage{graphicx,epsfig}
\usepackage{psfig}

\begin{document}

\markboth{L.~Jenkovszky, O.~Kuprash, J.~L\"ams\"a, and R.~Orava}
{Low-Mass Diffraction at the LHC}

%%%%%%%%%%%%%%%%%%%%% Publisher's Area please ignore %%%%%%%%%%%%%%
\catchline{}{}{}{}{}
%%%%%%%%%%%%%%%%%%%%%%%%%%%%%%%%%%%%%%%%%%%%%%%%%%%%%%%%%%%%%%%%%%%

\title{LOW-MASS DIFFRACTION AT THE LHC}

%\vskip 0.5cm \centerline{\bf Low-Mass Diffraction at the LHC}
%\vskip 0.3cm \centerline{ L.~Jenkovszky $^{a\diamond}$, O.~Kuprash
%$^{b\clubsuit}$, J.~L\"ams\"a $^{c\heartsuit}$, and R.~Orava
%$^{d,e\spadesuit}$}

\author{L.L~JENKOVSZKY}

\address{\sl {Bogolyubov Institute for Theoretical Physics (BITP)
of the \\National Academy of Sciences of Ukraine,\\
 14-b, Metrolohichna str., Kiev, 03680, Ukraine \\
\mbox{jenk@bitp.kiev.ua}
}}

\author{O.E.~KUPRASH}

\address{\sl {Faculty of Physics, Taras Shevchenko National University, \\
2, Glushkova ave., Kiev, 03022, Ukraine \\
\mbox{oleg.kuprash@desy.de}
}}

\author{J.W.~L\"AMS\"A}

\address{\sl {Physics Department, Iowa State University,\\
Ames, 50011 Iowa, USA \\
\mbox{jerry.lamsa@cern.ch}
}}

\author{R.~ORAVA}

\address{\sl {Division of Elementary Particle Physics, Helsinki Institute of Physics,\\
P.O. Box 64 (Gustaf H\"allstr\"ominkatu 2a), F1-00014 University of Helsinki, Finland;\\
CERN, CH-1211 Geneva 23, Switzerland \\
\mbox{rorava@cc.helsinki.fi; risto.orava@cern.ch}
}}

%\centerline{$^a$ \sl {BITP, Academy of Sciences of Ukraine, Kiev
%03680, Ukraine}} \centerline{$^b$ \sl {Taras Shevchenko National
%University, Kiev, Ukraine}} \centerline{$^c$ \sl {Physics
%Department, Iowa State University, Ames, 50011 Iowa, U.S.A.}}
% \centerline{$^d$ \sl {Helsinki
%Institute of Physics, Division of Elementary Particle Physics,}}
%\centerline{\ \ \sl {P.O. Box 64 (Gustaf H\"allstr\"ominkatu 2a),
%FI-00014 University of Helsinki, Finland}} \centerline{$^e$  \sl
%{CERN, CH-1211 Geneva 23, Switzerland}}

%\vskip 0.1cm

%\vfill

%% $
%% \begin{array}{ll}
%%  ^{\diamond}\mbox{{\it e-mail address:}} &
%%    \mbox{jenk@bitp.kiev.ua}\\
%% ^{\clubsuit}\mbox{{\it e-mail address:}} &
%%    \mbox{oleg.kuprash@desy.de}\\
%%    ^{\heartsuit}\mbox{{\it e-mail address:}} &
%%    \mbox{jerry.lamsa@cern.ch}\\
%%    ^{\spadesuit}\mbox{{\it e-mail address:}} &
%% \mbox{rorava@cc.helsinki.fi; risto.orava@cern.ch}
%% \end{array}
%% $

\maketitle
\pub{Received 30.06.2011}{Revised (Day Month Year)}

\begin{abstract}
The expected resonance structure for the
low-mass single diffractive states from a Regge-dual model
elaborated paper by the present authors in a previous is predicted. Estimates for
the observable low-mass single diffraction dissociation (SDD) cross sections and efficiencies
for single diffractive events
simulated by PYTHIA 6.2 as a function of the diffractive mass are given.

\end{abstract}

\ccode{PACS Nos.: 11.55.-m, 11.55.Jy, 12.40.Nn}

\section{Single Diffraction Dissociation (SDD)}
Low-mass single diffraction dissociation (SDD) will be among the first
measurements at the LHC \cite{review}. While high-mass diffractive
scattering \cite{Roy} \cite{Kaidalov} \cite{Dino1} \cite{Dino2} receives much attention -
mainly due to its relatively straightforward interpretation
through triple Regge formalism and successful measurements at the ISR,
HERA and Tevatron - the low-mass SD still lacks both experimental
measurement and theoretical understanding.

The unpolarized elastic $pp$ differential cross section is given
by \cite{model}

\begin{equation}\label{sigma}
    \frac{d\sigma}{dt}=\frac{[3\beta F^p(t)]^4}{4\pi\sin^2[\pi\alpha_P(t)/2]}(s/s_0)^{2\alpha_P(t)-2},
\end{equation}
%\begin{equation}\label{sigma}
%    \frac{d\sigma}{dt}=[3\beta F_1^p(t)]^4|(-is/s_0)^{\alpha_P(t)-1}|^2.
%\end{equation}
where the constant $\beta$ is obtained by normalizing
$d\sigma/dt\approx 80$ mb/GeV$^2$ at $t=0.$  The linear Pomeron
trajectory is $\alpha(t)=1.08+0.25t;$ the increase with energy is
provided by the supercritical Pomeron intercept. A dipole form can
be used for the form factor \cite{JL} \cite{DL}
\begin{equation}\label{FF1}
F_1^p(t)=\frac{4m^2-2.9t}{4m^2-t}\frac{1}{(1-t/0.71)^2},
\end{equation}
where $m$ denotes the proton mass. Assuming Regge factorization,
the double differential cross section for single diffractive
scattering (SD)
\begin{equation}\label{SD}
    pp\rightarrow pX,
\end{equation}
can be written as \cite{JL} \cite{DL}
\begin{equation}\label{DD1} \frac{d^2\sigma}{dtdM^2}\sim
\frac{9\beta^4[F_1^p(t)]^2}{4\pi\sin^2[\pi\alpha_P(t)]}(s/M^2)^{2\alpha_P(t)-2}
\Bigl[\frac{W_2}{2m}\Bigl(1-M^2/s\Bigr)-mW_1(t+2m^2)/s^2\Bigr],
\end{equation}
where $W_i,\ \ i=1,2$ are related to the structure functions of
the nucleon, and $W_2\gg W_1.$ At high $M^{2}$, the $W$s are
Regge-behaved, while at small $M^2$ their behavior is dominated by
nucleon resonances $N^*$s. Thus, the behavior of (\ref{DD1}) in
the low missing mass region largely depends on the transition form
factors or resonance structure functions, typically
$\gamma^*p\rightarrow N^*\rightarrow \pi p.$ The knowledge of the
inelastic form factors (or transition amplitudes) is crucial for
the calculation of low-mass diffraction dissociation (\ref{SD}) by
using equation (\ref{DD1}). The transition amplitudes (inelastic
form factors) were introduced in Ref. \cite{model}. We use a
supercritical Pomeron, with the intercept $\alpha_P(0)\approx 1.1$
and slope $0.2$ GeV$^{-2}$.
\begin{figure}
\begin{center}
\includegraphics[width=.60\textwidth]{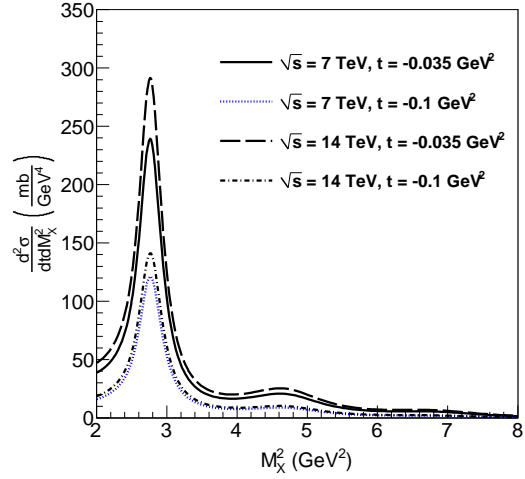}
\end{center}
\vspace{-1cm}
\caption{Cross section calculated for the single diffraction 
dissociation (SD) by using Eq. (\ref{DD1}) for the LHC energies
$\sqrt{s}= 7$ and $14$ TeV for different values of the 
four-momentum transfer, $-t=(p-p')^2$. 
For details see Ref. $^{5}$.  %\cite{model}.
 \protect\label{Fig:SD}} 
\end{figure}
At the LHC energies, terms $M^2/s$ and $(t+2m^2)/s^2$ can be
safely neglected in Eq. (\ref{DD1}). Furthermore, the signature
factor in the amplitude
$\frac{1-e^{-i\pi\alpha_P(t)}}{sin\pi\alpha_P(t)},$ used in Refs.
\cite{JL} \cite{DL} is replaced by an exponential (see Ref.
\cite{review}) $e^{-i\pi\alpha_P(t)/2}.$ For the elastic proton
form factor, $F^p(t),$ a dipole form
\begin{equation}\label{FF1}
F_1^p(t)=\frac{1}{(1-t/0.71)^2},
\end{equation}
is used.

At the LHC energies, Eq. (\ref{DD1})  now simplified as
\begin{equation}
\label{DD2} \frac{d^2\sigma}{dtdM_X^2}\approx
\frac{9\beta^4[F^p(t)]^2}{4\pi}(s/M_X^2)^{2\alpha_P(t)-2}
\frac{W_2}{2m}\,.
\end{equation}

In Fig. \ref{Fig:SD} double differential cross sections are shown
for two c.m.s. energies, $\sqrt{s}=7$ and $14$ TeV with different
values of the four-momentum squared, $t$.

\section{Forward detectors and efficiencies at the LHC}
In the following, an analysis of acceptance effects is carried out
for a predicted observed spectrum of low-mass SD cross section.
The forward detector  lay-out of the CMS experiment is used in
calculations. Somewhat similar results can be obtained by the
ATLAS experiment \cite{review}.
\begin{figure}[h]
\begin{center}
\includegraphics[width=.50\textwidth]{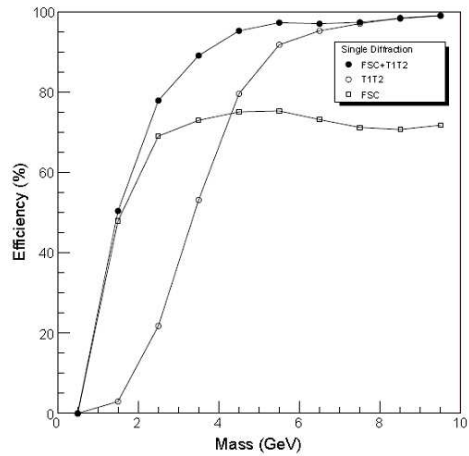}
\end{center}
\caption{The calculated efficiencies for single diffractive events
simulated by PYTHIA 6.2 as a function of the diffractive mass.
Five charged particles (hits) are required in any of the FSC, or
at least one track in the $\eta$ region covered by T1 or T2.}
\label{Eff}
\end{figure}
At the CMS intersect, IP5, the TOTEM T1 and T2 trackers can be
employed to detect SD masses below $M_X = 10$ GeV (see Fig. 2 and
Ref. \cite{review}).  The total $\eta$ regions spanned by T1 and
T2 is, approximately $3 < |\eta| < 7$.  The combination of T1 and
T2, labeled the $T1T2$ veto, can be used to reject any event
having a charged track in the T1T2 eta region. At the ATLAS
intersect, IP1, a similar $\eta$ region is, in principle, covered
 by the forward calorimetry and LUCID luminosity detector.

Forward Shower Counters, FSC, can be added closely surrounding the
beam pipes, at $60\ m < |z| < 85\ m$ (between the MBXW elements of
D1), and at further locations out to $z = \pm 140\ m$ on both sides
of the interaction point, IP5 (or similarly the experimental areas
of ALICE, ATLAS or LHCb), see Ref. \cite{Lamsa}.

The trigger efficiencies of the forward detector systems
\footnote{In our analysis we use the forward detector system of
the CMS/TOTEM experiment.} \cite{review} for single-diffractive
interactions, as a function of the diffractive mass, simulated by
PYTHIA and the GEANT program sequence are shown in Figure 2.

Three trigger possibilities for data collection are considered.
The first would be to trigger on events without any restriction
(minimum bias), i.e., {\it no veto}.  The second would be a
trigger with a veto on a given eta range, i.e., a T1T2 veto.  The
third would be a combination of T1T2 + FSC veto.

\section{Model calculations}

%%%%%%%%%%%%%%%%%%
%% \begin{figure}[h]
%% \parbox[r]{8.5cm}{\epsfxsize=55mm
%% \epsffile{1.eps}} %\hfill~
%% \parbox[c]{9.cm}{\epsfxsize=55mm
%% \epsffile{2.eps}}
%% \parbox[t]{8.cm}{\caption{Differential cross section $d\sigma/dM^2_{X}$,
%% corrected for CMS detectors acceptance. \label{fig:n1}}}\hfill~
%% \parbox[t]{8.cm}{\caption{Same as on the left panel, but
%% for different detector combinations \label{fig:n1width}}}
%% \end{figure}
%%%%%%%%%%%%%%%%%%

\begin{figure}[ht]
\begin{minipage}[b]{0.45\linewidth}
\centering
\includegraphics[scale=0.3]{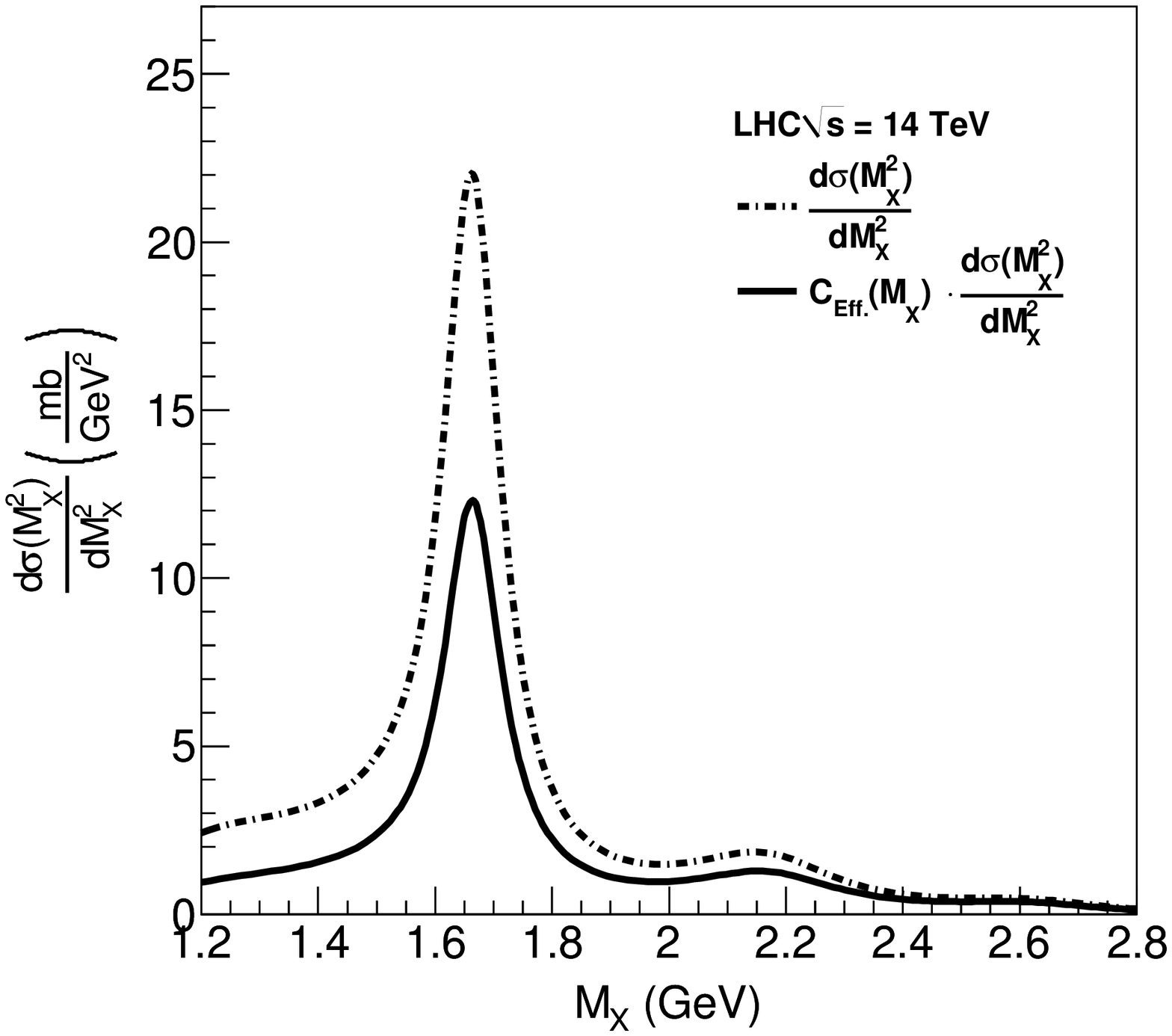}
\vspace{-1cm}
\caption{Differential cross section $d\sigma/dM^2_{X}$,
corrected for CMS detectors acceptance. \label{fig:n1}}
\end{minipage}
\hspace{0.5cm}
\begin{minipage}[b]{0.45\linewidth}
\centering
\includegraphics[scale=0.3]{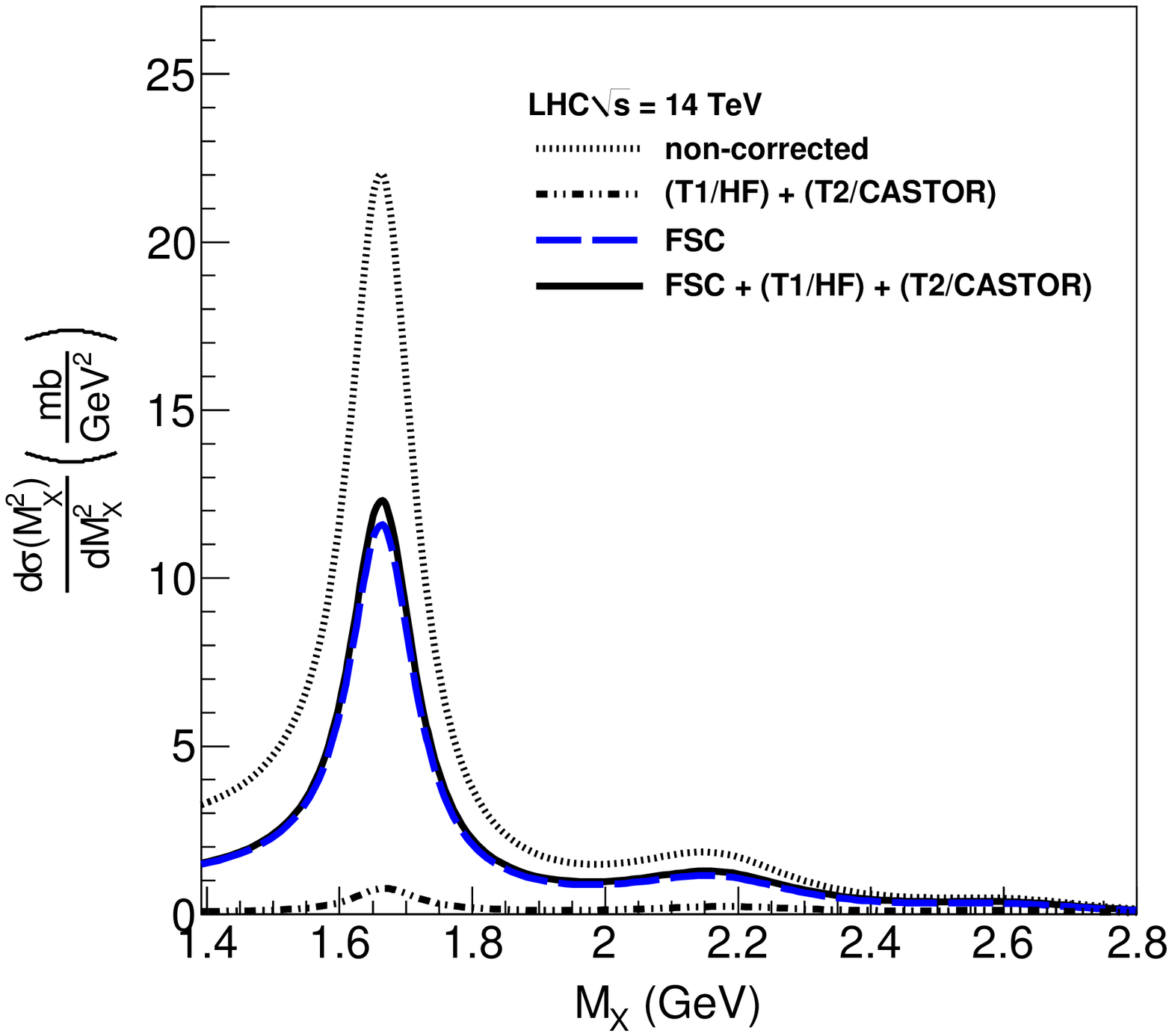}
\vspace{-1cm}
\caption{Same as on the left panel, but
for different detector combinations \label{fig:n1width}}
\end{minipage}
\end{figure}

\begin{figure}[ht]
\begin{minipage}[b]{0.45\linewidth}
\centering
\includegraphics[scale=0.3]{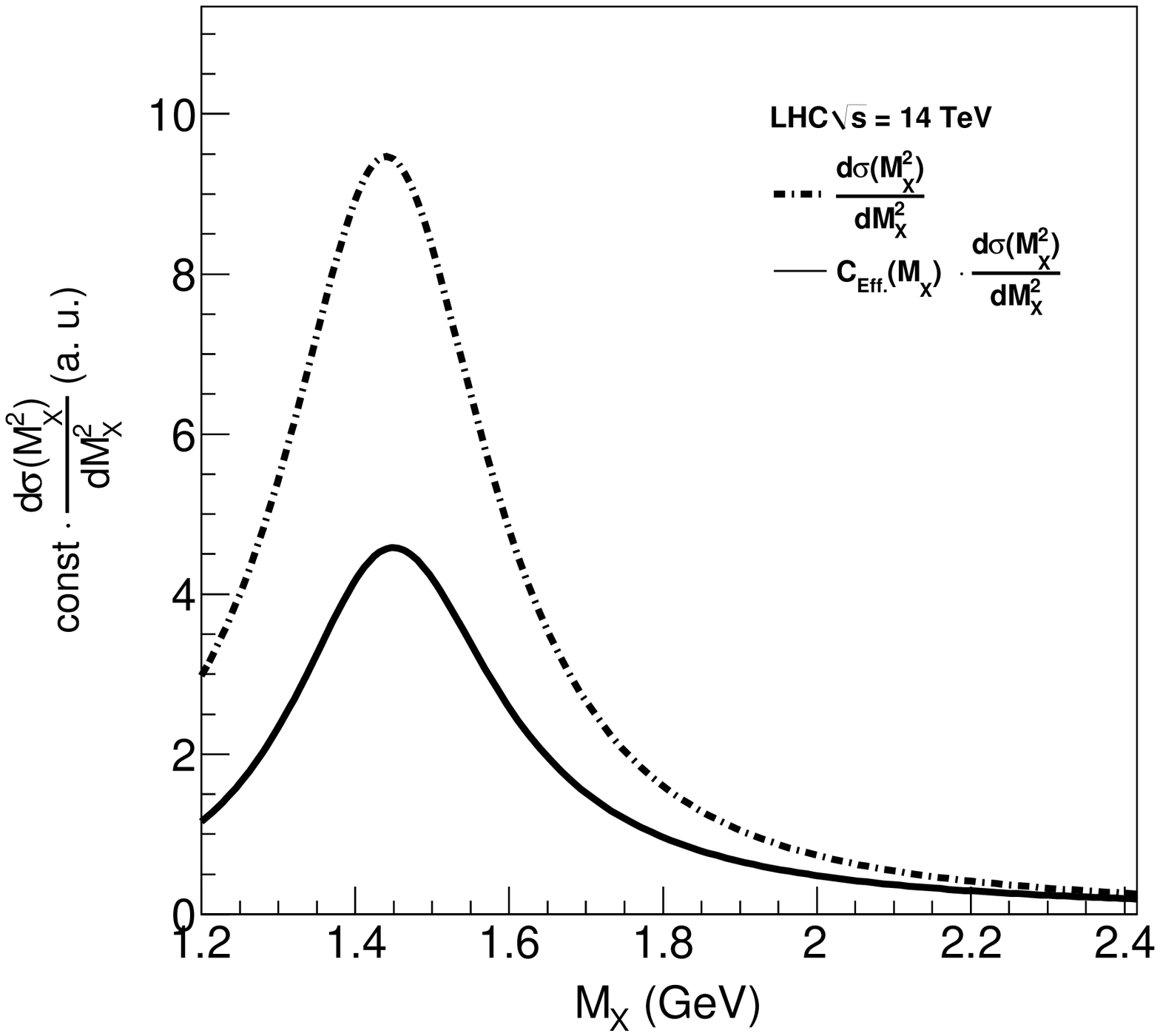}
\vspace{-1cm}
\caption{Roper resonance signal, corrected for CMS
detectors acceptance (arbitrary units). \label{fig:n1}}
%\vspace{1cm}
\end{minipage}
\hspace{0.5cm}
\begin{minipage}[b]{0.45\linewidth}
\centering
\includegraphics[scale=0.3]{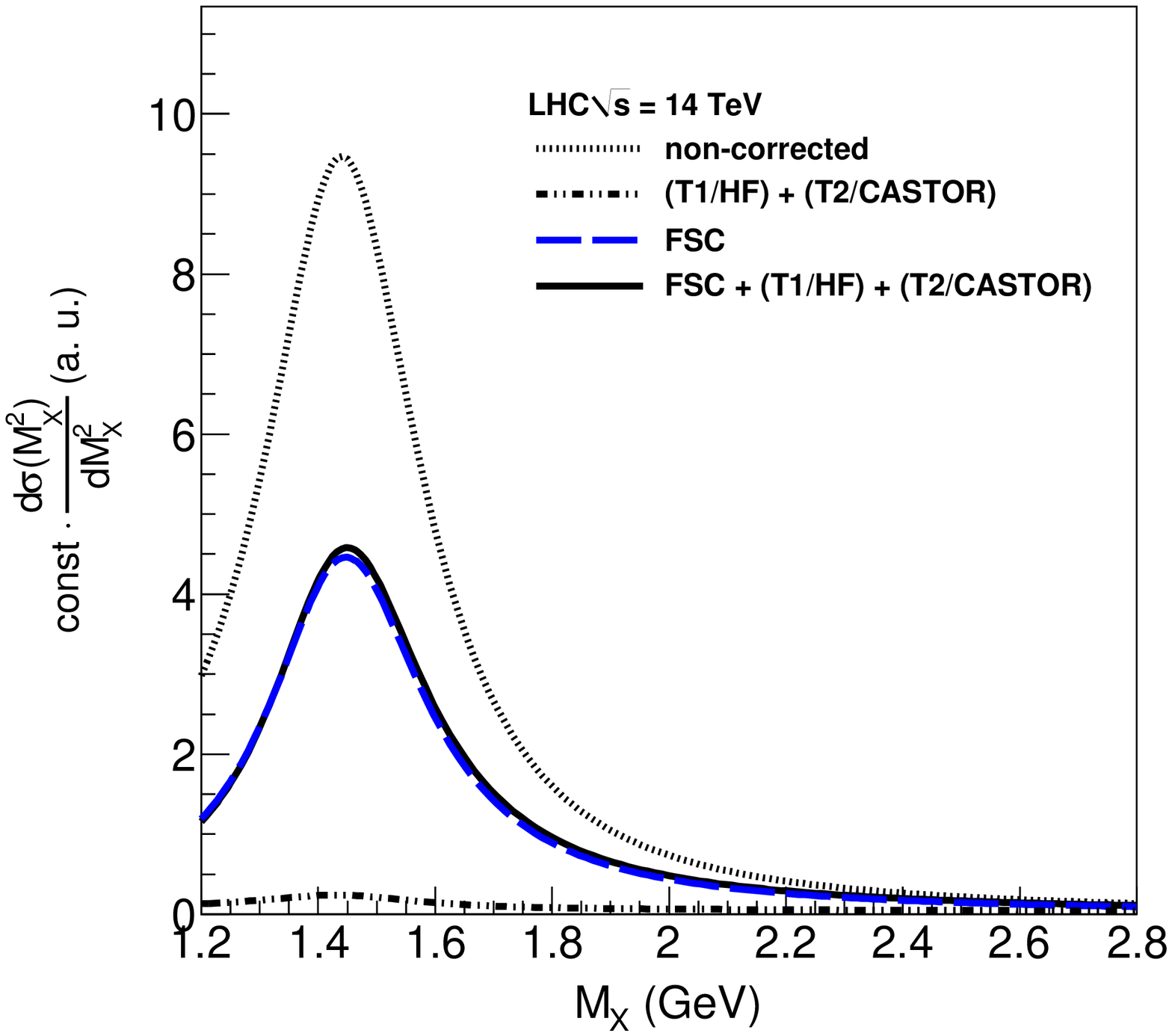}
\vspace{-1cm}
\caption{Same as on the left panel, but
for different detector combinations (arbitrary units).
%\vspace{1cm}
\label{fig:n1width}}
\end{minipage}
\end{figure}

%%%%%%%%%%%%%%%%%%
%% \begin{figure}[h]
%% \parbox[r]{8.5cm}{\epsfxsize=75mm
%% \epsffile{3bis.eps}} \hfill~\parbox[c]{9.cm}{\epsfxsize=75mm
%% \epsffile{4.eps}}
%% \parbox[t]{8.cm}{\caption{Roper resonance signal, corrected for CMS
%% detectors acceptance (arbitrary units). \label{fig:n1}}}
%% \hfill~\parbox[t]{8.cm}{\caption{Same as on the left panel, but
%% for different detector combinations (arbitrary units).
%%  \label{fig:n1width}}}
%% \end{figure}
%%%%%%%%%%%%%%%%%%

%%%%%%%%%%%%%%%%%%
%% \begin{figure}[h]
%% \parbox[r]{8.5cm}{\epsfxsize=75mm
%% \epsffile{5.eps}} \hfill~\parbox[c]{9.cm}{\epsfxsize=75mm
%% \epsffile{6.eps}}
%% \parbox[t]{8.cm}{\caption{Differential cross section $d\sigma/dy$,
%% corrected for CMS detectors acceptance (rapidity $y = -\ln \frac{M^2_X}{m_p \sqrt{s}}$). \label{fig:n1}}}
%% \hfill~\parbox[t]{8.cm}{\caption{Roper resonance signal as
%% function of rapidity, corrected for CMS
%% detectors acceptance (arbitrary units).
%%  \label{fig:n1width}}}
%% \end{figure}
%%%%%%%%%%%%%%%%%%

In Figures 3-6, the model calculations for the low mass SD process
(Eq. 6, Fig. 1) are corrected by the experimental efficiencies
(Fig. 2). The forward detector systems T1 and T2 (or equivalently
the HF and CASTOR calorimeters) facilitate detection of forward
diffractive masses down to about 4 GeV, far above the three
dominating $N^*$ states.

An approximate calculation of the diffractive mass can be made throught its
relation to the size of the rapidity gap adjacent to the scattered proton.
 The adjacent rapidity gap is defined as the gap between the diffractive
proton (close to the beam rapidity) and the nearest particle in rapidity.
The correspondence between the diffractive mass $M$ and the pseudorapidity gap
$\Delta \eta$ is $M_{X}/\sqrt{s} \sim e^{\Delta \eta}$.

\begin{figure}
\begin{center}
\includegraphics[width=0.6\textwidth]{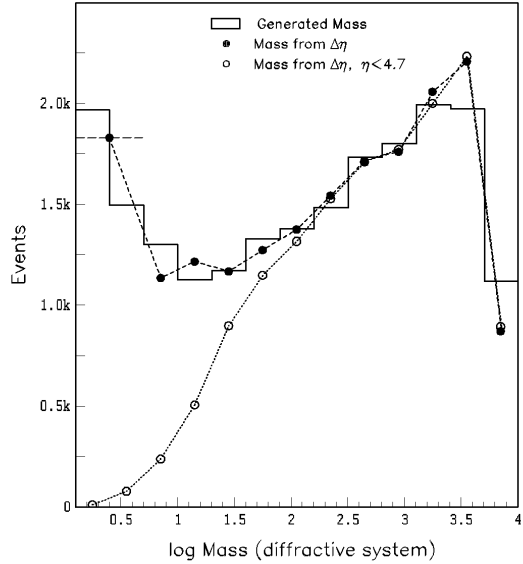}
\end{center}
\vspace{-0.5cm}
\caption{Generated diffractive mass and reconstructed diffractive mass calculated for two cases of detector coverage}
\label{Risto}
\end{figure}

To provide a more precise (although model dependent) measurement, the PYTHIA
program has been used to determine the correlation between the diffractive
mass and the size
of the rapidity gap.  Figure 7 shows the actual (generated) diffractive mass
together with that calculated by the above method, for two cases: (a) for
full eta coverage, and (b) for
limited eta range $|\eta| < 4.7$, i.e., the nominal CMS coverage.

\begin{figure}[ht]
\begin{minipage}[t]{0.45\linewidth}
\centering
%\vspace{1cm}
\includegraphics[scale=0.3]{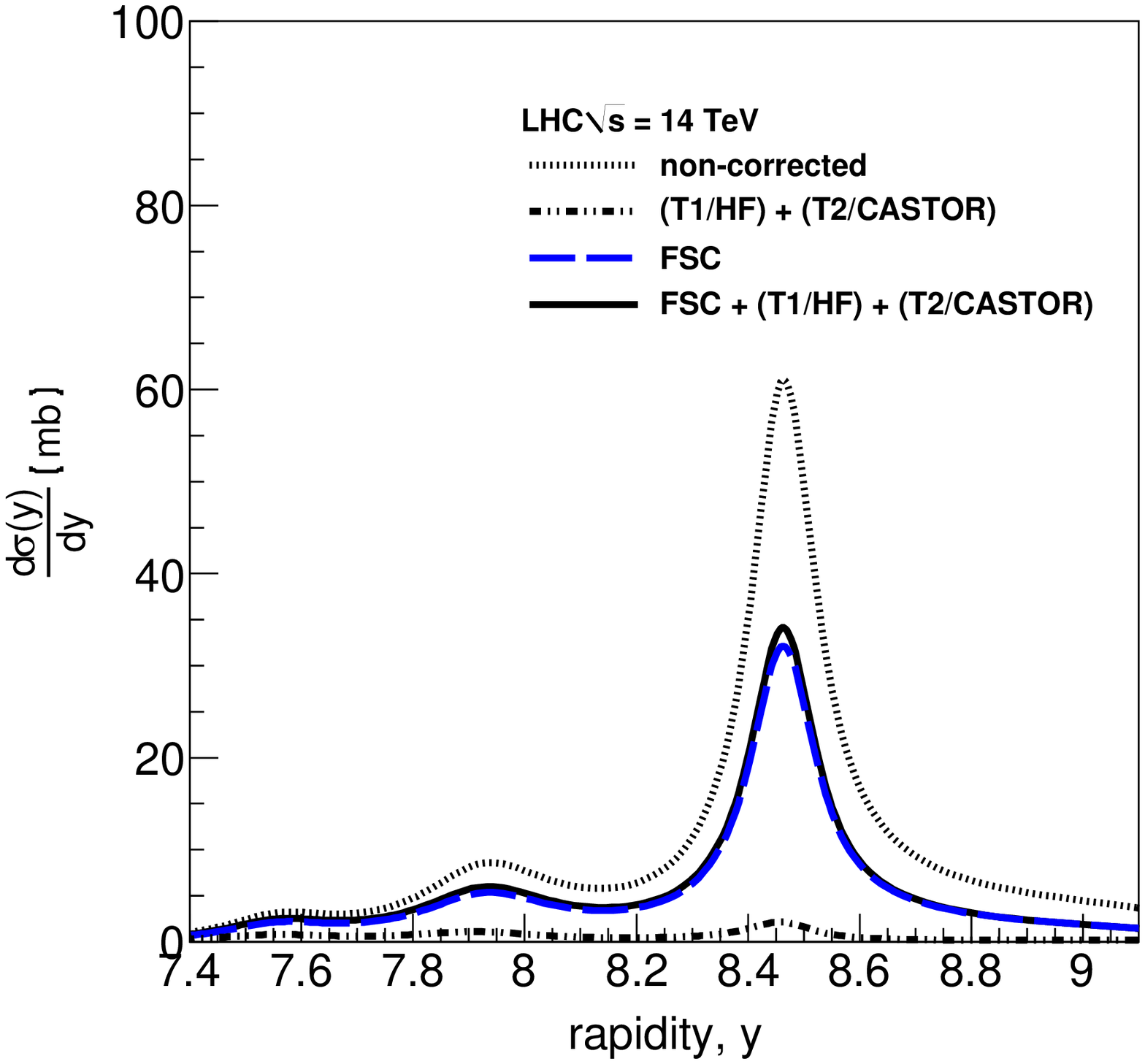}
\vspace{-1cm}
\caption{Differential cross section $d\sigma/dy$,
corrected for CMS detectors acceptance 
(rapidity $y = -\ln \frac{M^2_X}{m_p \sqrt{s}}$)
. \label{fig8}}
\end{minipage}
\hspace{0.5cm}
\begin{minipage}[t]{0.45\linewidth}
\centering
\includegraphics[scale=0.3]{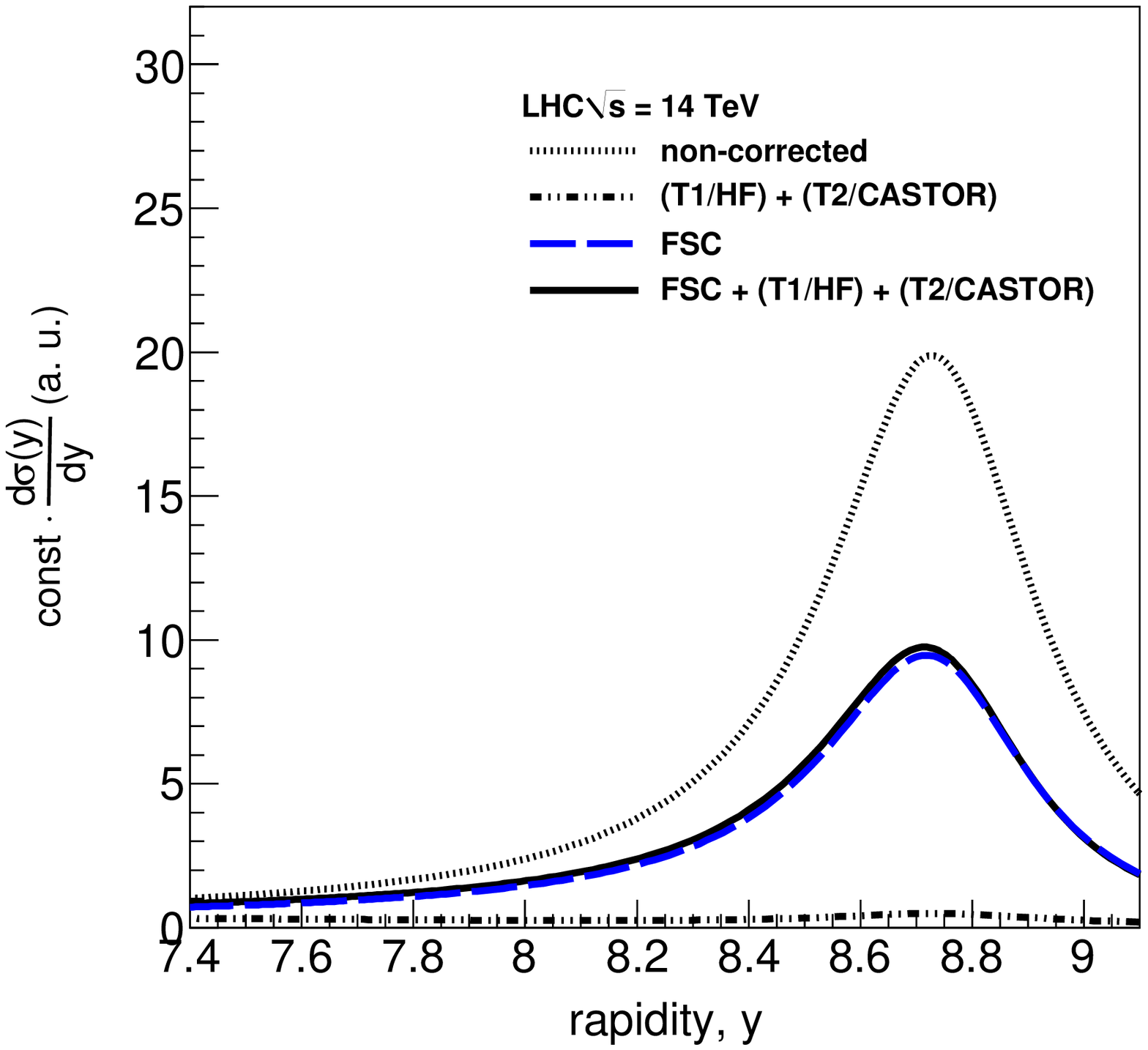}
\vspace{-1cm}
\caption{Roper resonance signal as
function of rapidity, corrected for CMS
detectors acceptance (arbitrary units).
 \label{fig9}}
\end{minipage}
\end{figure}

The efficiency of the Forward Shower Counters for detecting forward
diffractive systems is high.  For low-mass single diffraction one relies
largely on the FSC's.  Measurement of the content (off-line) of individual
FSC counters, which cover different eta-ranges, provides more differential
tests of the diffractive event simulation.  From the various rates, with
knowledge of the FSC efficiencies, the background contributions can be
estimated and subtracted from different situations (e.g., different $M_{X}$).
 Correlations between the FSC counters can be determined and compared with
expectations. These will be used to make more precise determinations of the
mass of the diffractive system.  The ultimate uncertainties to be achieved
will come from work which is still in progress.
Another valuable check will be the independence of all the measured cross
sections on the instantaneous luminosity.

In the model presented here, there is no significant contribution
from the so called Roper resonance, $N^*(1440)$. For completeness,
the case where the $N^*(1440)$ dominates the low SD masses was
considered. This mass region would be efficiently covered by the
additional Forward Shower Counters (see Fig.4, fsc references),
and $x$ of the $N^*(1440)$ final states would be covered.

\section{Conclusions}

%The low mass single diffractive (SD) processes remain largely
%unknown at current collider energies. In this paper model
%calculations together with simulated experimental efficiencies are
%presented for the single diffractive proton-proton events below
%$M_x = 10$ GeV. The $N^*$ states $N^*(1675,�$) dominate the low
%mass region and do not allow simple $1/M_X^2$
% extrapolations to be used for
%estimating the event rates missed by the base line  experiments.
%On the basis of our model calculations, the relative rate and
%uncertainty of  low mass SD events seen by the forward detector
%arrangements at the LHC (CMS/TOTEM is used for this exercise but
%similar conclusions could be drawn for ATLAS) can be evaluated.

The low mass single di®ractive (SD) processes remain largely unknown at current collider
energies. In this paper model calculations together with simulated experimental efficiencies
are presented for the single diffractive proton-proton events below $M_x = 10$ GeV. The $N^*$
states $N^*(1675,\; 1680, ...)$ MeV dominate the low mass region and do not allow simple $1/M^2_X$
extrapolations to be used for estimating the event rates missed by the base line experiments.

As further steps comparison of the event rates and uncertainties below $M_X = 10$ GeV should be made, including:

$\bullet\ 1/M_X^2$
 extrapolation to 1 GeV vs. the N* spectrum calculation,

$\bullet$  uncertainties in these.

Although low-mass diffraction at the ISR and Tevatron energies was studied in quite a number of papers,
the details of the complicated resonance structure in the missing
mass at the LHC energies still leaves quite a number of open problems. For example, the slope of the diffraction cone is known
to increase monotonically as the missing mass increases beyond the resonance region,
see, e.g. Fig. 19 of Ref. \cite{Dino1}, this may not be the case in the resonance region, scrutinized in the present paper,
see Fig 1. A behavior of the local slope (as function of the missing mass and $t$) in the resonance region
may affect considerably the efficiency of the forward shower counters. A complementary means to study
low-mass diffraction are finite-energy mass rules (FMSR), relating low- and high-missing mass dynamics (see Ref. \cite{Dino2}.
It should be remembered, however, that FMSR contain information only on the average, i.e. the integrated behavior of the
resonant amplitude, without providing details about separate resonances contributions. Moreover, the resonance contribution 
in the FMSR integral should be appended by an independent elastic contribution and a vaguely known background, 
see Ref. \cite{Otranto}. We intend to come back to these interesting and important problems in a future study.
%}

On the basis of our model calculations, the three dominating $N^*$ states remain below
the detection thresholds of the current forward detectors at the LHC. By installing Forward
Shower Counters (see Refs. above), the rates of these small mass diffractive events can be
recorded thus facilitating an accurate measurement of the total pp cross section at the LHC.

\section*{Acknowledgements}

%We thank V.~Magas for discussions. RO gratefully acknowledges the
%Academy of Finland for support. O.K. is grateful to Rainer
%Schicker and the EU program "Diffractive and Electromagnetic
%Processes at the LHC" for their support.
We thank V. Magas for discussions and K. Goulianos for useful correspondence.  RO gratefully acknowledges
the Academy of Finland for support. O.K. is grateful to Rainer Schicker and the EU program ``Diffractive and
Electromagnetic Processes at the LHC'' for their support. L.J. was supported by the Project
``Matter under Extreme Conditions'' of the Nat. Ac. Sc. of Ukraine, Dept. of Astronomy
and Physics.

\end{document}